# An Interactive Transactive Energy Mechanism Integrating Grid Operators, Aggregators and Prosumers

Peng Hou, Guangya Yang, Junjie Hu, Philip J. Douglass, *and* Yusheng Xue

*Abstract*— With the decreasing cost of solar photovoltaics (PV) and battery storage systems, more and more prosumers appear in the distribution systems. Accompanying with it is the trend of using home energy management systems (HEMS). HEMS technologies can help the households to schedule their energy prosumption with aims such as reduced electricity bills or increased self-sufficiency. However, their economic-driven operation can affect the grid security. Therefore, it is paramount to design a framework that can accommodate the interests of the key stakeholders in distribution systems, namely the grid operators, aggregators, and prosumers. In this paper, a novel transactive energy based operational framework is proposed. On the upper level, aggregators will interact with distribution grid operators through transactive approach to ensure the grid interests are satisfied. If there are grid issues, the aggregator will interact with the prosumers through a designed price adder. The simulation results indicate that the proposed framework can effectively accommodate the prosumers operation in distribution systems while keeping the key stakeholders' interests.

## NOMENCLATURE

**Acronym**
| | |
|---|---|
| PVST | PV and battery storage system |
| DER | Distributed energy resources |
| DR | Demand response |
| HEMS | Home-energy management system |
| ND | Normalized difference |
| CoS | Cost of security |

**Parameter**
| | |
|---|---|
| $i, t, k, r, j$ | Respectively represent the index of prosumer, Time slot, aggregator, rolling optimization process, and bus no. |
| $A_{i,k}$ | The objective function of $i^{th}$ prosumer associated with the $k^{th}$ aggregator |
| $B$ | The objective function of the aggregator when transactive energy approach is triggered |
| $\mu_{t,k}^{Agg}$ [€/kWh] | Forecasted retail electricity price at hour $t$ of prosumer associated with aggregator $k$ |
| $\mu_{t,i}^{Sell}/\mu_{t,i}^{buy}$ [€/kWh] | Price of selling/purchasing 1 kWh electricity by prosumer $i$ at hour $t$ |
| $\mu_t^{Up}/\mu_t^{Down}$ [€/kWh] | Up-regulation/down-regulation price at hour $t$ |
| $\mu_t^{TSO}, \mu_t^{DSO}$ [€/kWh] | Grid tariff at hour $t$ from TSO/DSO |
| $\mu^{ETax}$ [€/kWh] | Electricity tax |
| $\mu_t^{DAM}$ | Day-ahead market price at hour $t$ |
| $N_{Agg}$ | Total number of aggregators in the system |
| $N_{bus}$ | Total number of buses in the studied distribution system. |
| $\omega_{buy}/\omega_{sell}$ [%] | Profit coefficient of aggregator for purchasing/selling respectively |
| $VAT$ [%] | Value Added Tax |
| $M_{DSO}$ [€/kWh] | Participation factor that represents the participation preference of each EV prosumer |
| $U^{max}, U^{min}$ | Minimum and maximum voltage limit in p.u. |
| $U^0$ | Initial voltages of the buses in the network in p.u |
| $P_{trans}^{Max}$ | Power capacity of the transformer, in p.u. |
| $P_i^+, P_i^-$ [kW] | Maximum charging/discharging power of prosumer $i$ |
| $P_{t,i}^{pv}$ [kW] | PV output of $i^{th}$ prosumer at hour $t$ |
| $P_{t,i}^{load}$ [kW] | Load consumption of $i^{th}$ prosumer at hour $t$ |
| $J_{21}^{-1}$ | V/P submatrix of the inverse Jacobian. |
| $c_{bat}, c_{Bd}$ [€/kWh] | Battery capital and degradation cost |
| $LET$ [year] | Battery life energy throughput |
| $L_c, L_s$ [kWh], $DoD$ | Cyclic lifetime, battery capacity and depth-of-discharge respectively |
| $\Omega_j$ | Set of bus no. |
| $T$ [hour] | Whole time window of the optimization horizon |
| $E_{s,i}$ [kWh] | Energy storage system capacity of $i^{th}$ prosumer |
| $\eta_{Ch}/\eta_{Dis}$ | Charging/Discharging efficiency of battery respectively |
| $SOC_{initial}, SOC^{max}, SOC^{min}$ | Initial SOC, maximum SOC, minimum SOC |
| $p$ | Iteration index for price adder |
| $\rho$ | Step size |
| $\epsilon$ | Convergence accuracy |

**Variable**
| | |
|---|---|
| $P_t^{Ch}/P_t^{Dis}$ [kW] | Charged/discharged power in the battery at hour $t$ |
| $P_{t,i}^B/P_{t,i}^S$ [kW] | Purchased/sold power by prosumer $i$ at hour $t$ |
| $P_{t,j}^{DSO}$ [kW] | Power schedule submitted to DSO at hour $t$ of bus No. $j$ |
| $P_{t,j}^{Agg}$ [kW] | Power schedule submitted to aggregator at hour $t$ of bus No. $j$ |
| $SOC$ | State of charge |
| $\delta_{t,i}^a/\delta_{t,i}^b$ [kW] | Binary indicator of exporting/importing energy to the grid by prosumer $i$ at hour $t$ |
| $\lambda$ [€/kWh] | Price adder |
| $\lambda^{rev}$ [€/kWh] | The revised price signal that broadcasted to prosumers |

This work was supported by the "Energy Technology Development and Demonstration Program under Grant EUDP7-1: 12551."
P. Hou is with the SEWPG European Innovation Center, Inge Lehmanns Gade 10, Aarhus C, 8000, Denmark (e-mail: houpeng@shanghai-electric.com).
G. Yang is with the Center of Electrical Engineering, Technical University of Denmark, Lyngby, 2800 Denmark (e-mail: gyy@elektro.dtu.dk).
P.J. Douglass is with Danish Energy, Vodroffvej 59, 1900 Frederiksberg C Denmark (e-mail: pdo@danskenergi.dk).
J. Hu is with School of Electrical and Electronic Engineering, North China Electric Power University, China. (e-mail: junjiehu@ncepu.edu.cn)
Y. Xue is with State Grid Electric Power Research Institute, Nanjing, China. (e-mail: xueyusheng@sgepri.sgcc.com.cn)

2## I. INTRODUCTION

The proliferation of DERs results in a paradigm change in the operation of the power system. Instead of purchasing energy produced by large generating companies through energy suppliers, more and more residents equip their homes with PV and battery storage to improve their self-sufficiency via HEMS [1]. However, the uncertainties in both the PV generation and the residents' preferences challenge the safe operation of the distribution system. To alleviate potential congestion and voltage violation problems, exploitation of the flexibility embedded in the DERs is a promising solution and thus draws worldwide attention.

Transactive energy (TE) represents a group of promising technologies that facilitate the operation and coordination among intelligent devices as well as stakeholders using value as the sole media. As only value is exchanged among the parties, their privacy and autonomy can be retained [2]. So far, various research works have been developed under this principle. In [3], transactions are expected to be made between the houses with HEMS and a corresponding transactive node while multiple objectives are considered. TE was also utilized to provide DR to the grid via the commercial building HVAC systems considering agent bidding strategies [4]. The DERs were coordinated in the form of virtual power plant (VPP) in [5] and optimally controlled in both day-ahead (DA') and real-time (RT) market. The whole problem was formulated as a two-stage optimization problem and the TE was applied in the outer layer to minimize the imbalance cost in the RT market. An agent-based TE architecture was proposed in [6], the distributed energy storage systems and various types of DR loads in different microgrids were controlled by a comprehensive energy management system. The simulation results showed that the proposed framework could effectively deal with the energy imbalances within the microgrids and lower the dependency of the microgrids on the main system. TE was also adopted in [7] for coordinating the generation and consumption in a rural off-grid microgrid. A comparison study of the performance of various TEs was presented in [8].

Though TE has been demonstrated to be effective for enabling transactions for DERs in the above works, the network constraints have not been taken into account fully. Thus, some works started to see the possibility of meeting the network constraints by TE [9][10]. In [9], both the distributed demand and supply were controlled indirectly via the TE. Under the assumption that the aggregators have the right to control the associated DERs when there is a requirement of meeting the system constraints, TE was applied in [10] for optimizing the EV prosumers' schedules. In [11], a non-bidding TE market was proposed which enabled the transactions between the distributed prosumers (transactive agents) and the distributed system operator (DSO). Under the TE, the real-time market participation of the thermostatically controlled load consumers was enabled in [12], the proposed strategy was demonstrated to be less sensitive to the forecast error since only the mean and volatility of clearing price over a future time window is required. To address the uncertainties in both the renewable energy production and market prices, a stochastic programming model was proposed in [13] where the TE approach was utilized to facilitate the operation of the rural micro-grids with the purpose of energy balancing. The TE was adopted in [14] for transactions among the neighboring microgrids; the simulation shows that the proposed method reduces the levelised cost of energy. A TE framework proposed in [15] regards the transactive agent as the aggregated demand.

The previous TE work for distribution system operation was addressing only interactions between mainly two parties, or multiple identical parties under certain context, while lacks a complete interaction framework covering the demand side, aggregators, and the distribution network operators. In this paper, we propose a framework to facilitate the operations of all three parties based on TE principle, with the aim of accommodating the prosumers' operation. TE is applied for activating the demand response from prosumers while meeting the distribution system's constraints. Specifically, the contributions include: 1) The role of aggregator is redefined compared with [4][6][10][16] (where the aggregator is assumed to have the right of controlling the DERs directly) and specified in this work as not only an energy retailer but also an aggregated ancillary service representative; 2) A new mixed-integer linear programming (MILP) model is proposed for PVST prosumer scheduling considering the difference in the price of selling and purchasing electricity. 3) A price signal is formulated addressing the interests of DSO, aggregators, and prosumers, while the responsiveness of prosumers is fully accounted. 4) A new TE approach is presented that can achieve a close-loop control strategy to ensure the operational goal. Based on this, the scalability of TE is significantly increased and the willingness of prosumer is truly respected.

The paper is organized as follows: The HEMS-based TE architecture is specified in the next section. Based on this, the optimization model is then described in section III. The simulation model and results are discussed in section 4. Conclusions are presented last.

## II. PROPOSED TRANSACTIVE ENERGY FRAMEWORK

The proposed scheme can be well integrated with many markets in the world. The section first discusses the tariff policy followed by the proposed TE framework.

### A. Current Tariff Model

A tariff model can be defined according to the customer's production or consumption and the connected voltage level, where the tariff that each type of customer should pay varies, e.g. based on waterfall principle [17]. Currently in many countries in the world there is no production tariff for the residential customers and residents pay only tariffs for each kWh of consumption as well as a subscription fee [17]. Additionally, time-of-use tariff is adopted by some DSOs in order to reduce the peak load [18]. The peak load tariff applies during the period between 17:00 – 20:00 from October 1st to March 31st. An example of Danish case is shown in Table I.

TABLE I. PRICE FOR ELECTRICITY TRANSPORT [EURO/kWH]

|     | Peak | Off-peak |
|-----|------|----------|
| TSO | 0.01 | 0.01 |
| DSO | 0.09 | 0.03 |

*B. Proposed Transactive Energy Network for Prosumer*

Existing work for demand side activation was primarily based on a fictitious price quantity response curve of consumers, where the energy consumption was expected to follow an external price signal based on the curve. However, it is hard to approximate such a curve in reality, and under the new privacy regulation, such information may not be available for electricity retailers and the grid operators. Instead, with increased use of controllable household solar PV and storage units and online appliances, prosumers tend to schedule their energy presumption to maximize their self-sufficiency and hedge the volatile external supply conditions. HEMS can help the prosumers to coordinate and optimize their schedules with respect to their forecasted production and consumption preferences. Such function is already adopted in commercial HEMS type of systems [19]. In this way, the residents' energy presumption information is not shared, hereby the privacy issues are resolved. There have been articles in the field favoring this model [20].

Prosumers equipped with HEMS may participate in the electricity market directly, provided that they have access to the marketplace and knowledge of operation. Else, they can subscribe to an electricity aggregator/retailor whose role is to guarantee their power supply as well as aggregate the distributed resources for market operation. In the latter context, HEMS communicates directly with the aggregator to submit their schedule and obtain price information based on their type of supply contracts. Such two-way communication among prosumers and aggregators enables a TE network. While from the grid operator side, to ensure efficient grid operation and fulfill the network security, a TE framework can be set between the aggregator and the network operator through a distribution independent system operator (DISO) [16]. In this way, a TE framework can be established to ensure a coordinated operation of prosumers, aggregators, and distribution grid operators at the same time.

In this framework, it is assumed there is a variable price contract between the aggregator and the prosumer, where the prosumer will schedule their presumption based on the predicted electricity price from the associated aggregator and the forecasted consumption and PV production. The objective of the HEMS is to control the prosumer's devices to minimize the electricity bills (or maximize profits in the case of excess PV generation). The schedule from the prosumers in each time interval will be collected by the aggregators for trade in the market according to their bidding strategy. The schedules of all the aggregators will be sent to the DSO to validate network security. If no congestion or voltage constraints violations occur, the DSO will accept all the schedules. Otherwise, a TE will be triggered between the DSO and the aggregators. A social welfare problem between DSO and aggregators will be resolved. To guarantee the fairness of the interaction [16], a third party DISO is defined to facilitate the transactive procedure between the aggregators and the DSO. The aggregator may bid in the balancing market to compensate for the derived imbalance cost due to TE. The transactive procedure will generate a price quantity representing the added cost of security to the aggregators due to network constraints. In literature, this price quantity has not been effectively used to activate the individual prosumers through a framework, hence the aggregators are assumed to bear the cost of security. In this work, instead, this price quantity is used to formulate a price adder (PA) and communicated to the prosumers through HEMS, which represents the network cost of electricity due to their schedule. The HEMS will add this PA to the original forecasted electricity price and reschedule the consumption and submit to the aggregator. If the new schedule still results in violations of network constraints, the whole process will repeat until security criteria are met. This PA, in fact, is equivalent to network tariff representing the network conditions in the coming time period.

The proposed TE strategy extends the previous work in several respects. Firstly, TE models proposed in literature usually consider only two parties, e.g. aggregator and prosumer, or aggregator and DSO, where to the authors' knowledge, there have not been work addressing the interactions of all three parties. Secondly, an innovative PA is designed to represent the network conditions and used to activate the response from the prosumers. Thirdly, the proposed TE framework will not stop unless the constraints are satisfied, which provides a close-loop framework for the prosumers operation in the distribution network. The overall conceptual architecture is shown in Fig. 1.

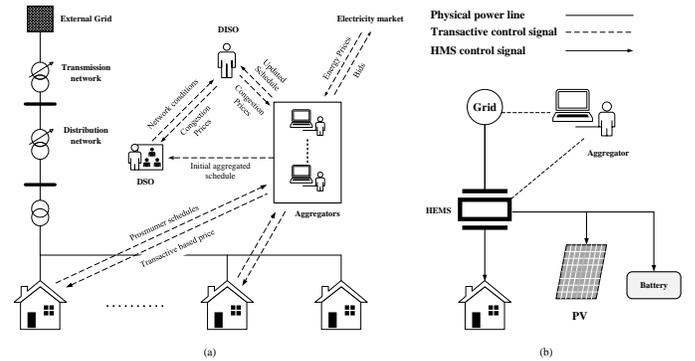

Fig. 1. (a) The proposed TE architecture. (b) HEMS for PVST prosumers.

*C. Rolling Operation for Prosumers*

The operation of prosumers is expected to follow a forward rolling time window, where within the time window the forecasted production and consumption is used for scheduling. The time window length is designed 3 hours, with 1-hour resolution. The proposed rolling window optimization (RWO) model is illustrated in Fig. 2.

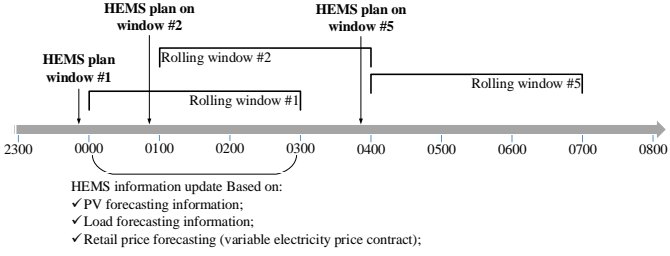

Fig. 2. Rolling procedure for HEMS controlled PVST prosumers.

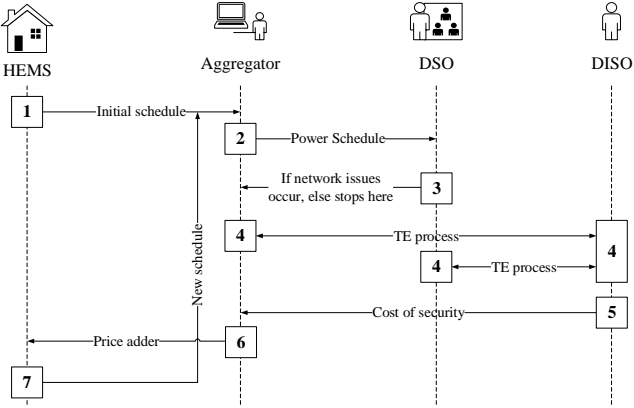

Fig. 3. The operation flow for the TE system in each RWO process.

The overall operation flow is shown in Fig. 3. It is an iterative process until the aggregated schedule from prosumers is valid for the network. In each iteration, the PVST prosumers will first optimize their own schedules according to the predicted information and submit the schedule to the subscribed aggregator. The aggregated schedules will be passed to DSO by the aggregators. If network constraints are violated, the TE mechanism will be triggered to help reschedule the PVST prosumer's schedule until the agreement reached among the aggregators and DSO (step 4). Afterward, the aggregator sends the PA to its associated prosumers and then obtains an updated schedule from them. The HEMS manages each prosumer's schedule and the control strategy is designed according to each PVST prosumer's preferences. The aggregator can indirectly influence the prosumer's behavior through a price adder.

## III. Optimization Models

The overall optimization problem is formulated as a social welfare maximization (and cost minimization) problem.

### A. Prosumer's Operation Model

For the PVST type of prosumers, HEMS mainly controls the storage unit to achieve economic operation. Since there are periods where there are surplus PV output and the energy purchasing and selling price from aggregator is asymmetric due tariffs and taxes, binary variables are required to indicate both the prosumer's action (buy/sell) as well as the battery operation status (charging/discharging). The specification of binary variables is included in Table II.

TABLE II. Specification of Binary and Bi-linear Terms

| Variable | Type | Definitions | Description |
|---|---|---|---|
| $\delta^1_{t,i}$ | Binary | | Exporting (sell) energy to the grid |
| $\delta^2_{t,i}$ | Binary | | Importing (buy) energy to the grid |
| $\delta^3_{t,i}, \delta^4_{t,i}$ | Binary | $\delta^1_{t,i}(P^{pv}_{t,i} - P^{load}_{t,i} - \delta^3_{t,i}P^{Ch}_{t,i} + \delta^4_{t,i}P^{Di}_{t,i}) \geq 0$ | The battery is in charging/discharging mode when it is exporting to the grid |
| $\delta^5_{t,i}, \delta^6_{t,i}$ | Binary | $\delta^2_{t,i}(P^{pv}_{t,i} - P^{load}_{t,i} - \delta^5_{t,i}P^{Ch}_{t,i} + \delta^6_{t,i}P^{Di}_{t,i}) \leq 0$ | The battery is in charging/discharging mode when the prosumer is importing from the grid |
| $z^1_{t,i}$ | Real > 0 | $z^1_{t,i} = \delta^1_{t,i}\delta^3_{t,i}P^{Ch}_{t,i}$ | The energy flows into the battery when the prosumer is charging while exporting to the grid; |
| $z^2_{t,i}$ | Real > 0 | $z^2_{t,i} = \delta^1_{t,i}\delta^4_{t,i}P^{Dis}_{t,i}$ | The energy flows out from the battery when prosumer is discharging and exporting to the grid |
| $z^3_{t,i}$ | Real > 0 | $z^3_{t,i} = \delta^2_{t,i}\delta^5_{t,i}P^{Ch}_{t,i}$ | The energy flows into battery when charging while the prosumer is importing from the grid |
| $z^4_{t,i}$ | Real > 0 | $z^4_{t,i} = \delta^2_{t,i}\delta^6_{t,i}P^{Dis}_{t,i}$ | The energy flows out from the battery when discharging while importing from the grid |

Considering the electricity products provided by the aggregators, the PVST prosumers' objective function (A) in each rolling procedure can be expressed in the following.

$$\text{Obj.:} \quad \max_{P^S, P^B} A_{i,k} = \sum_{t \in T} \left[ P^S_{t,i} \mu^{Sell}_{t,k} + P^B_{t,i} \mu^{buy}_{t,k} \right] \quad (1)$$

$$\mu^{buy}_{t,k} = (1 + VAT)\left(\mu^{Agg}_{t,k} + \mu^{TSO}_t + \mu^{DSO}_t + \mu^{ETax}_t\right) \quad (2)$$

$$\text{S.t.:} \quad P^S_{t,i} = \delta^1_{t,i}\left(P^{pv}_{t,i} - P^{load}_{t,i} - \delta^3_{t,i}P^{Ch}_{t,i} + \delta^4_{t,i}P^{Dis}_{t,i}\right) \geq 0 \quad (3)$$

$$P^B_{t,i} = \delta^2_{t,i}\left(P^{pv}_{t,i} - P^{load}_{t,i} - \delta^5_{t,i}P^{Ch}_{t,i} + \delta^6_{t,i}P^{Dis}_{t,i}\right) \leq 0 \quad (4)$$

$$0 \leq P^{Ch}_{t,i} \leq P^+_i \quad (5)$$

$$0 \leq P^{Dis}_{t,i} \leq P^-_i \quad (6)$$

$$SOC^{min}_i \leq SOC_{t,i} \leq SOC^{max}_i \quad (7)$$

$$SOC_{1,i} = SOC_{initial,i} \quad (8)$$

$$\delta^1_{t,i} + \delta^2_{t,i} = 1 \quad (9)$$

$$\delta^3_{t,i} + \delta^4_{t,i} = 1 \quad (10)$$

$$\delta^5_{t,i} + \delta^6_{t,i} \leq 1 \quad (11)$$

$$\begin{cases} SOC_{t+1,i} = SOC_{t,i} + \dfrac{\left(\delta^1_{t,i}\delta^3_{t,i} + \delta^2_{t,i}\delta^4_{t,i}\right)P^{Ch}_{t,i}\eta_{Ch,i}}{E_{s,i}} \\ \qquad - \dfrac{\left(\delta^1_{t,i}\delta^4_{t,i} + \delta^2_{t,i}\delta^6_{t,i}\right)P^{Dis}_{t,i}}{E_{s,i}\eta_{Dis,i}} \end{cases} \quad (12)$$

It should be noticed that the above optimization problem contains bi-linear terms in eq. (3), (4) and (12), which make the problem non-convex. This intractable problem can be resolved by linearizing the bi-linear term through introducing extra variables $z^1_{t,i}$ to $z^4_{t,i}$ as listed in Table II. This is the so-called big M technology [21]. By using big M technology, the optimization model can then be reformulated. For simplicity, the index $i$ is neglected in the following. The full mathematical formulation via big M is specified in the supporting document.

The energy schedule from the PVST prosumers for time interval $t$ at bus $j$, $P_{t,j}^{Agg}$, will be submitted to the DSO via aggregators. If the schedule violates the system constraints, the TE mechanism will be triggered; otherwise, this schedule will be accepted. A simple battery life cost model is adopted as proposed in [22], mathematically it is expressed as:

$$c_{Bd} = c_{bat}/L_{ET} \quad (13)$$

$$L_{ET} = L_c L_s DoD \quad (14)$$

To consider the impact of the BDDC on the scheduling, the objective function (1) is modified into the following,

Obj.: $$\min_{P^{sell},P^{buy}} A_i = \sum_{t\in T} -\left[P_t^{sell}\mu_t^{Sell} + P_t^{buy}\mu_t^{buy} - c_{Bd}\left(z_t^2 + z_t^4\right)/\eta_{Dis}\right] \quad (15)$$

### B. DSO's Objective

The responsibility of the DSO is to meet the energy demand of each aggregator while ensuring the overall operational schedule meets the distribution system constraints. The DSO's optimization problem can be written as follows:

$$\min_{P^{DSO}} D = \sum_{j\in N_{bus}}\sum_{t\in T}\mu_{DSO}(P_{t,j}^{DSO} - P_{t,j}^{Agg})^2 \quad (16)$$

$$P_{t,j}^{Agg} = \sum_{i\in\Omega_j}\left(P_{t,i}^S + P_{t,i}^B\right) \quad (17)$$

S.t. $$-P_{trans}^{Max} \leq \sum_{j\in N_{bus}} P_{t,j}^{DSO} \leq P_{trans}^{Max} \quad (18)$$

$$U_{trans}^{min} \leq U_{t,j}^0 - J_{21}^{-1}P_{t,j}^{DSO} \leq U_{trans}^{max} \quad (19)$$

The corresponding prosumers' schedules of bus j will be aggregated by aggregator as shown in (17). A sensitivity method ignoring reactive power variation is adopted to calculate the grid voltage in (19) according to [10].

### C. Aggregators Operation Model

The aggregators' operation is to satisfy the demand from market participation. Here, we consider aggregator has already a purchase agreement from the day-ahead market. The schedule submitted from the prosumers represents a modification of the day-ahead schedule. The deviation from the initial schedule may introduce extra cost to aggregator. To minimize the cost, this modification can be considered as balancing power provided to the grid. Each aggregator's optimization problem in this phase can be expressed as follows:

Obj.: $$\min_{P^S,P^B} B = \sum_{t\in T}\left(P_{t,i}^B\mu_t^{Down} - P_{t,i}^S\mu_t^{Up}\right) \quad (20)$$

$$\delta_{t,i}^{a1} + \delta_{t,i}^{a2} \leq 1 \quad (21)$$

$$0 \leq P_{t,i}^S \leq 2P_i^-\eta_{Dis,i}\delta_{t,i}^{a1} \quad (22)$$

$$0 \leq P_{t,i}^B \leq 2P_i^+\eta_{Ch,i}\delta_{t,i}^{a2} \quad (23)$$

The schedule sent to DSO can be expressed as,

$$\sum_{k\in N_{Agg}}\sum_{i\in\Omega_j}\left(P_{t,i}^B - P_{t,i}^S\right)_k = P_{t,j}^{DSO} \quad (24)$$

The aggregator can either provide up regulation or down regulation which is indicated in (21). The range of the up/down regulation energy is described in (22) and (23) respectively. Considering the difficulties in obtaining SOC information of PVST prosumers via smart meters in the real-world, a modification is made in this work which relaxes the upper boundaries of charging/discharging limits by assuming a full battery storage in t-1 time interval. Eq. (24) shows the common interests between aggregator and DSO.

### D. Social Welfare Maximization Between DSO and Aggregators

A social welfare optimization problem can be formulated as,

Obj.: $$\min_{P_{t,i,k}^S,P_{t,i,k}^B,P_{t,j}^{DSO}} \sum_{k\in N_{Agg}}\sum_{t\in T}\sum_{i\in\Omega_j} B_k\left(P_{t,i,k}^S,P_{t,i,k}^B\right) + \sum_{j\in N_{bus}}\sum_{t\in T} D\left(P_{t,j}^{DSO}\right) \quad (25)$$

s.t. (18)-(19), (21)-(24).

It can be seen that (24) is a shared constraint for the optimization problem of DSO and the aggregators.

### E. Transactive Energy Model via ADMM

Eq. (25) will be solved iteratively between the DSO model and the aggregator's model through a price signal which represents the cost of security (CoS). The model is non-convex due to the presence of binary variables. ADMM is adopted here to solve the problem [23]. Firstly, an augmented Lagrangian of the problem (25) is written in the following.

$$L_p\left(P_{t,j}^{Agg*},P_{t,j}^{DSO},\lambda_{t,j}\right) = \sum_{k\in N_{Agg}}\sum_{t\in T}\sum_{i\in\Omega_j} B_k\left(P_{t,i,k}^S,P_{t,i,k}^B\right) + \sum_{j\in N_{bus}}\sum_{t\in T_{RO}} D\left(P_{t,j}^{DSO}\right) + \sum_{j\in N_{bus}}\sum_{t\in T_{RO}}\lambda_{t,j}\left[P_{t,j}^{Agg*} - P_{t,j}^{DSO}\right] + \frac{\rho}{2}\left\|\sum_{j\in N_{bus}}\sum_{t\in T_{RO}}\left[P_{t,j}^{Agg*} - P_{t,j}^{DSO}\right]\right\|_2^2 \quad (26)$$

Where $\rho>0$. To solve (26), the ADMM includes the iterations in the following.

$$P_{t,j}^{Agg,p+1} := \arg\min_{P_{i,j}^{Agg}} L_p\left(P_{t,j}^{Agg*},P_{t,j}^{DSO,p},\lambda_{t,j}^p\right) \quad (27)$$

$$P_{t,j}^{DSO,p+1} := \arg\min_{P_{i,j}^{DSO}} L_p\left(P_{t,j}^{Agg*,p+1},P_{t,j}^{DSO},\lambda_{t,j}^p\right) \quad (28)$$

$$\lambda_{t,j}^{p+1} := \lambda_{t,j}^p + \rho\left(P_{t,j}^{Agg*,p+1} - P_{t,j}^{DSO,p+1}\right) \quad (29)$$

A constant step size $\rho$ is utilized to update $\lambda$ which is defined as 0.8 in this study. The convergence criterion is described in the following.

$$\left|\lambda_{t,j}^{p+1} - \lambda_{t,j}^p\right| \leq \varepsilon \quad (30)$$

where $\varepsilon$ is set as 0.005 in this study. After solving (26), the new schedule that is accepted by both DSO and aggregators can be formulated in the following.

$$P_{t,j}^{Agg*} = P_{t,j}^{Agg} + P_{t,j}^B - P_{t,j}^S \quad (31)$$

### F. Rescheduling of PVST Prosumers via a PA

The price quantity reflecting CoS should be communicated to the prosumers to reflect the network conditions, as it is induced due to the schedule of the prosumers. This has not been addressed in literature since the converged price can be much higher than normal electricity retail price due to rigid technical constraints that makes it no practical ground to be directly applied. In this work, we proposed a method to apply this price quantity to formulate a PA. First of all, the difference before and after initial schedule is normalized and represented by ND,

$$ND_{t,j} = \left(P_{t,j}^{Agg} - P_{t,j}^{Agg*}\right) \Big/ \max_{t\in[1,T], j\in\Omega_j}\left(P_{t,j}^{Agg} - P_{t,j}^{Agg*}\right) \quad (32)$$

There are totally four states when the initial schedules of PVST prosumers are required to be updated which has been concluded in Table III.

TABLE III . PA DIRECTION ACCORDING TO THE SIGN OF ND AND CoS

| ND | CoS | PA |
| --- | --- | --- |
| + | + | + |
|   | − | + |
| − | + | − |
|   | − | − |

In Table III, '+' means positive value and vice versa. A positive ND means that the agreed schedule between DSO and aggregator requires the PVST prosumers to use less energy from the grid or inject more energy to the grid. In such situation, it can be imagined that no matter the sign of CoS, a positive PA should be given to the prosumer so that the right response can be activated from PVST prosumer. In contrast, if ND has a negative sign, a negative PA should be applied. Finally, the actualized PA $\lambda_{t,j}^{rev}$ is formulated by the following formula

$$\lambda_{t,j}^{rev} = \left|\lambda_{t,j}^{TE}\right| ND_{r,j} \quad (33)$$

Upon receiving the PA from the aggregator, each PVST prosumer will reschedule their presumption by adding the PA to their forecasted price,

Obj.: $$\min_{P^S, P^B} A_{i,k} = \sum_{t\in T} -\left[\left(P_t^S + \lambda_t^{rev}\right)\mu_{t,k}^{Sell} + \left(P_t^B + \lambda_t^{rev}\right)\mu_{t,k}^{buy} - c_{Bd}\left(z_t^2 + z_t^4\right)\eta_{Dis}^{-1}\right] \quad (34)$$

The process of section III. A to III. F will be repeated until the schedule is accepted by the DSO.

### G. Discussion

In order to have the scheme work in reality, the following conditions need to be fulfilled,

- Prosumers can schedule their presumption with a certain accuracy level and are willing to respond to external prices;
- The marginal cost of the TE operation is low, which means the communication channels are reliable, and mathematically the problems can be solved quickly and reliably;
- A balance or imbalance market is available for the aggregators to modify their day-ahead schedule;

## IV. CASE STUDY

A representative low voltage grid is used to illustrate the efficacy of the framework.

### A. Parameter Settings

In this work, it is assumed that there are in total 18 PVST prosumers contracted with two different aggregators in a 0.4 kV low voltage distribution system which is the same test system in [16]. The power transformer capacity allocated to all residents in that area is 220 kW. The minimum and maximum voltage of the bus is set at 0.9 and 1.1 p.u. respectively. The parameters of batteries are specified in Table IV.

TABLE IV. BATTERY PARAMETERS

| Type | $E_b$ (kWh) | $SOC_{min}$ (%) | $SOC_{max}$ (%) | $P^+/P^-$ (kW) | $\eta_{Ch}/\eta_{Dis}$ (%) | $C_{bd}$ (Euro/kwh) |
| --- | --- | --- | --- | --- | --- | --- |
| 1 | 13.1 | 20 | 90 | 2.86 | 0.9/0.95 | 0.07 |
| 2 | 25.4 | 20 | 85 | 5.57 | 0.9/0.95 | 0.07 |
| 3 | 21.8 | 20 | 85 | 4.77 | 0.9/0.95 | 0.07 |
| 4 | 12.3 | 20 | 90 | 2.70 | 0.9/0.95 | 0.07 |
| 5 | 12.8 | 20 | 85 | 2.81 | 0.9/0.95 | 0.07 |

### B. Results and Discussion

Due to low diversity in PV production and likely similar forecast of the electricity price, schedules from prosumers can have high coincidence where there can be periods network constraints are violated. To simplify the study, it is assumed in the case study that all the prosumers under one aggregator have the same forecasting of electricity prices. The predicted variable electricity price that applied for the PVST prosumers are compared with the original variable electricity price in Fig. 4.

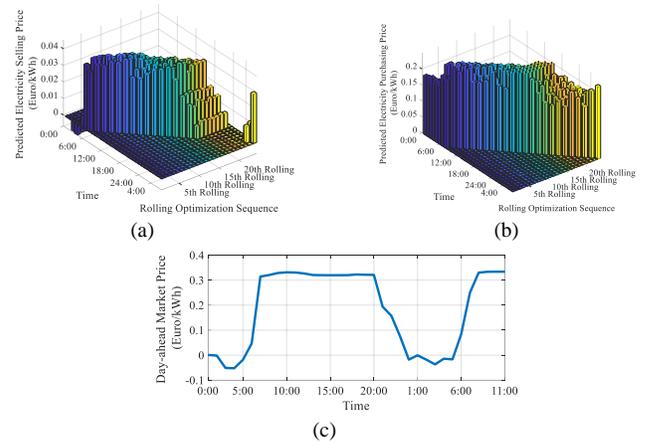

Fig. 4. Variable electricity price provided by the aggregator. (a) The predicted electricity selling price in each rolling process (prosumers in Aggregator 1). (b) The predicted electricity purchasing price in each rolling process (prosumers in Aggregator 1). (c) The day-ahead market price.

In this work, the rolling window length is 8 hours. The day-ahead market prices shown in Fig. 4(c) is obtained in [24], between 00:00 to 23:00, 5 Mar 2019 for DK2 area. The profits of each aggregator are considered by using a profit coefficient, which is a profit margin that each aggregator expects assuming each aggregator is a price-taker. This profit coefficient concept applies in both purchasing and selling prices of the aggregator and reflects the differences in aggregators' bidding strategies. The predicted purchasing/selling electricity prices by the PVST prosumers associated with the aggregator 1 are shown in Fig. 4(a) and (b) which is obtained based on the prices in Fig. 4(c).

To be concise, the predicted electricity prices from aggregator 2 is neglected. The coefficients are included in Table V.

TABLE V. PROFIT COEFFICIENT SPECIFICATION

| Aggregator No. | $\omega_{buy,t}$ (%) | $\omega_{sell,t}$ (%) |
|---|---|---|
| 1 | 10 | 8 |
| 2 | 12 | 11 |

It can be seen in Table V that the aggregator 1 gives cheaper electricity product to its contracted customers while purchases the surplus energy from the customers with lower price as well compared with the business strategy from aggregator 2. According to Table V, the energy purchasing/selling price can be expressed in the following.

$$\mu_{t,k}^{Agg} = \left(1+\omega_{buy,k}\right)\mu_t^{DAM} \tag{35}$$

$$\mu_{t,k}^{sell} = \left(1+\omega_{sell,k}\right)\mu_t^{DAM} \tag{36}$$

Based on the procedure specified in section III, the simulation results of all the prosumers are shown below.

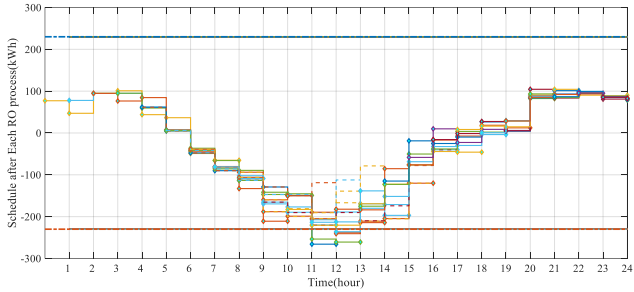
(a). Aggregated schedule of prosumers before/after TE in each RWO process.

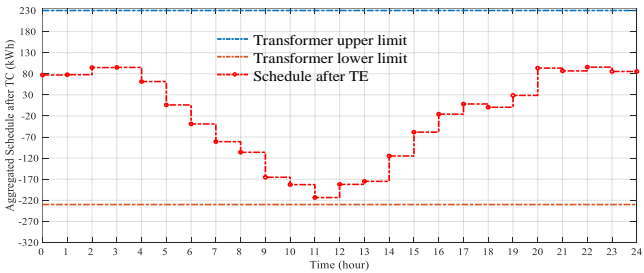
(b). Aggregated schedule that is finally submitted to DSO.

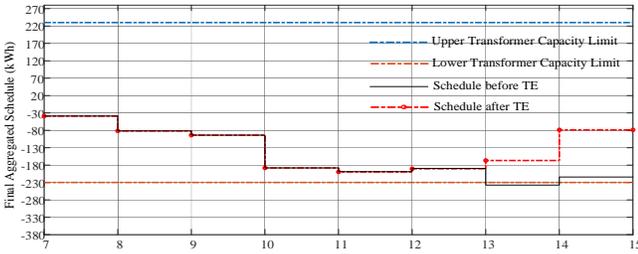
(c). Aggregated schedule of prosumers before/after TE in 6th RWO process.

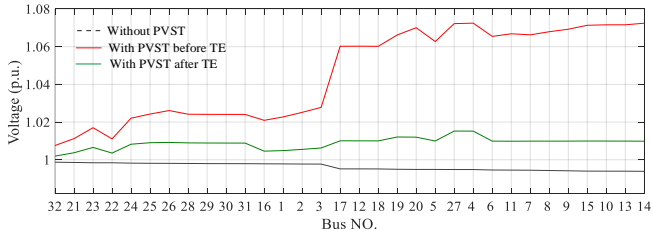
(d). System voltage before/after TE at 14th time interval in 6th RWO process.

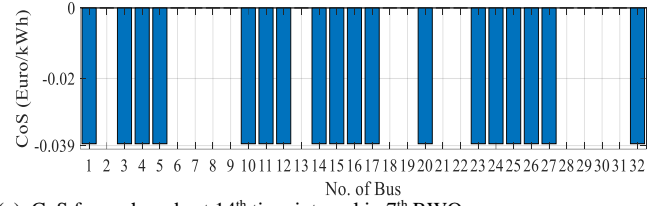
(e). CoS for each node at 14th time interval in 7th RWO process.

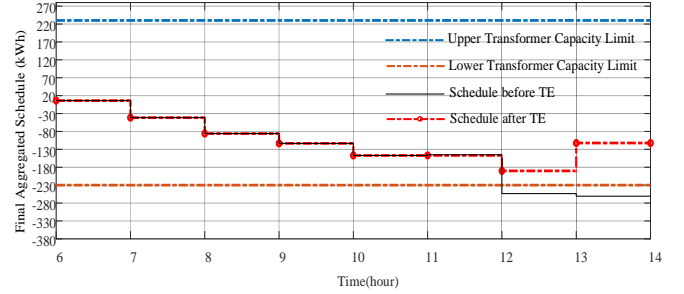
(f). Aggregated schedule of prosumers before/after TE in 7th RWO process.

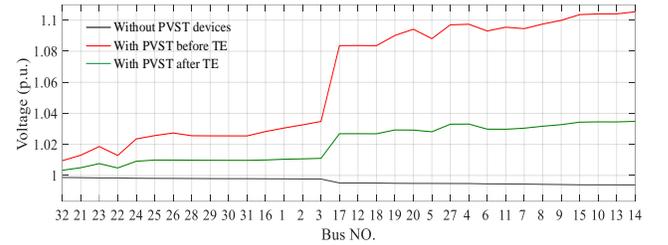
(g). Grid voltage before/after TE at 14th time interval in 7th RWO process.

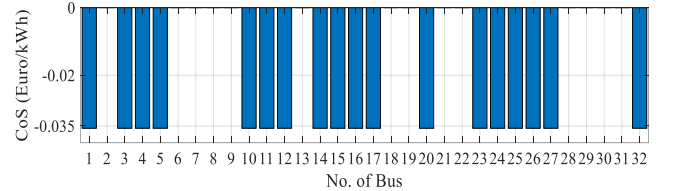
(h). CoS for each node at 14th time interval in 7th RWO process.

Fig. 5. Summary of TE results.

The aggregated schedule before and after the TE in each RWO process is compared in Fig. 5(a) where the solid color lines represent the schedules for each RWO without TE process and the dot lines represent schedules for each RWO with TE. The final aggregated schedule after each RWO is shown in Fig. 5(b). Due to the congestion constraint, the TE will be activated at 14th time interval. The aggregated schedules before/after TE in 6th rolling process are shown in Fig. 5(c) while the voltage profile is shown in Fig. 5(d). It can be seen that there is a remarkable voltage rise in each bus node. Correspondingly, the CoS is shown in Fig. 5(e). The aggregated schedules before/after TE in 7th rolling process are shown in Fig. 5(f) while the voltage profile is shown in Fig. 5(g). Both congestion and voltage violation problems occur in this period. After TE, the system voltage constraints are met as indicated by the green line in Fig. 5(g) while the congestion problem is also solved as shown in Fig. 5(c). The corresponding CoS is illustrated in Fig. 5(h). The CoS for each node is the same even when the voltage violation problem is solved. This benefits from the flexibility of PVST prosumers especially the storage units.

To elaborate on this, we reduce the battery size of the PVST prosumer on Bus No. 16 and 32 to 13.1 kWh and 12.3 kWh to 4.8 kWh and 3.6 kWh, power rating to 1.05 kW and 0.80 kW respectively and rerun the program. The overall simulation results are quite similar to the figures illustrated in Fig. 5. However, the CoS for $7^{th}$ rolling process shows different feature which is shown in Fig. 6.

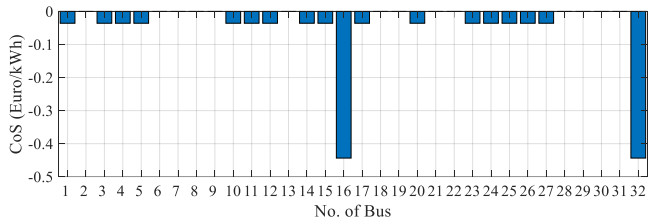

Fig. 6. CoS for each node at $14^{th}$ time interval in $7^{th}$ RWO process when assuming smaller batteries.

Due to the voltage constraints, larger CoS was induced at Bus No. 16 and 32 respectively. In this case, the PVST prosumers with smaller batteries cannot help the grid when they are asked to reschedule, thus higher prices are paid to so that they will have the willingness to change their schedule. This is not fair for those customers, as the network voltage issues are affected by all customers. In practice, we recommend broadcasting the same PA signal for all the customers.

## V. CONCLUSION

In this paper, a new TE framework for distribution systems with prosumers is proposed. The framework contains two interactions, aggregators with DSOs, and aggregators with prosumers. A PA is formulated to reflect the cost of security for prosumer's operation. This PA is, in fact, a dynamic tariff component imposed by the network conditions. The proposed new TE structure can preserve the privacy information of PVST prosumers when participating TE market to help meet the system constraints. The procedure has close-loop characteristics that guarantee the response from the prosumers.

In practice, the flexibility of the prosumers may not be sufficient to mitigate the grid issues. In this case, the stopping criteria of TE may be relaxed to a certain extent while there is still room for DSOs to use other control method to regulate the voltage and congestion issues. Alternatively, the DSO optimization problem in the TE framework can be extended to a distribution system optimal power flow problem, where more control variables can be introduced.

REFERENCES

[1] J. von Appen, T. Stez, M. Braun, and A. Schmiegel, "Local Voltage Control Strategies for PV Storage Systems in Distribution Grids," *IEEE Trans. Smart Grid,* vol. 5, no. 2, pp. 1002-1009, Feb. 2014.
[2] Z. Liu, Q. Wu, M. Shahidehpour, et. al, "Transactive Real-time Electric Vehicle Charging Management for Commercial Buildings with PV On-site Generation," *IEEE Trans. Smart Grid*, vol. 10, no. 5, pp. 4939-4950, Sept. 2018.
[3] H. Wu, A. Pratt, and S. Chakraborty, "Stochastic optimal scheduling of residential appliances with renewable energy sources," *IEEE Power Energy Soc. Gen. Meet.*, Denver, CO, 26-30 Jul. 2015.
[4] H. Hao, C. D. Corbin, K. Kalsi, and R. G. Pratt, "Transactive Control of Commercial Buildings for Demand Response," *IEEE Trans. Power Syst.*, vol. 32, no. 1, pp. 774–783, Apr. 2016.
[5] J. Qiu, K. Meng, Y. Zheng, and Z. Y. Dong, "Optimal scheduling of distributed energy resources as a virtual power plant in a transactive energy framework," *IET Gener. Transm. Distrib.*, vol. 11, no. 13, pp. 3417–3427, Oct. 2017.
[6] H. S. V. S. K. Nunna, and D. Srinivasan, "Multiagent-Based Transactive Energy Framework for Distribution Systems with Smart Microgrids," *IEEE Trans. Ind. Informatics*, vol. 13, no. 5, pp. 2241–2250, Mar. 2017.
[7] G. Prinsloo, A. Mammoli, and R. Dobson, "Customer domain supply and load coordination: A case for smart villages and transactive control in rural off-grid microgrids," *Energy*, vol. 135, pp. 430–441, Sept. 2017.
[8] J. Lian, H. Ren, Y. Sun, and D. Hammerstrom, "Performance Evaluation for Transactive Energy Systems using Double-auction Market," *IEEE Trans. Power Syst.*, vol. 34, no. 5, pp. 4128-4137, Oct. 2018.
[9] P. H. Divshali, B. J. Choi, and H. Liang, "Multi-agent transactive energy management system considering high levels of renewable energy source and electric vehicles," *IET Gener. Transm. Distrib.*, vol. 11, no. 15, pp. 3713–3721, Nov. 2017.
[10] J. Hu, G. Yang, C. Ziras, and K. Kok, "Aggregator Operation in the Balancing Market Through Network-Constrained Transactive Energy," *IEEE Trans. Power Syst.*, vol. 34, no. 5, pp. 4071-4080, Oct. 2018.
[11] S. Moazeni, and B. Defourny, "Distribution system controls assessment in a nonbinding transactive energy market," *2017 North Am. Power Symp. NAPS 2017*, Sept. 2017.
[12] S. Behboodi, D. P. Chassin, N. Djilali, and C. Crawford, "Transactive control of fast-acting demand response based on thermostatic loads in real-time retail electricity markets," *Appl. Energy*, vol. 210, pp. 1310–1320, Jan. 2018.
[13] M. Daneshvar, M. Pesaran, and B. Mohammadi-ivatloo, "Transactive energy integration in future smart rural network electrification," *J. Clean. Prod.*, vol. 190, pp. 645–654, Jul. 2018.
[14] S. A. Janko and N. G. Johnson, "Scalable multi-agent microgrid negotiations for a transactive energy market," *Appl. Energy*, vol. 229, pp. 715–727, Nov. 2018.
[15] J. Li, C. Zhang, Z. Xu, J. Wang, J. Zhao, and Y. J. Zhang, "Distributed Transactive Energy Trading Framework in Distribution Networks," *IEEE Trans. Power Syst.*, vol. 33, no. 6, pp. 7215–7227, Jul. 2018.
[16] J. Hu, G. Yang, H. W. Bindner, and Y. Xue, "Application of Network-Constrained Transactive Control to Electric Vehicle Charging for Secure Grid Operation," *IEEE Trans. Sustain. Energy*, vol. 8, no. 2, pp. 505–515, Sept. 2016.
[17] L. Aaberg, "Country specific issues related to DSO tariffs." [Online]. http://www.nordicenergyregulators.org/wp-content/uploads/2017/02/DSO-tariffs-in-Denmark.pdf
[18] R. de S. Ferreira, L. A. Barroso, P. R. Lino, M. M. Carvalho, P. Valenzuela, "Time-of-Use Tariff Design Under Uncertainty in Price-Elasticities of Electricity Demand: A Stochastic Optimization Approach," *IEEE Trans. on Smart Grid*, vol. 4, no. 4, pp. 2285-2295, Apr.2013.
[19] "SMA sunny home 2.0." [Online]. https://www.europe-solarstore.com/sma-sunny-home-manager-2-0.html
[20] P. Siano, G. De Marco, A. Rolán, V. Loia, "A Survey and Evaluation of the Potentials of Distributed Ledger Technology for Peer-to-Peer Transactive Energy Exchanges in Local Energy Markets," *IEEE Systems Journal*, vol. 13, no. 3, pp. 3454-3466, Mar. 2019.
[21] D. Mignone, "The REALLY BIG Collection of Logic Propositions and Linear Inequalities," Feb. 2002.
[22] R. C. Leou, "Optimal Charging/Discharging Control for Electric Vehicles Considering Power System Constraints and Operation Costs," *IEEE Trans. Power Syst.*, vol. 31, no. 3, pp. 1854–1860, Jul. 2016.
[23] S. Boyd, N. Parikh, E. Chu, et. al, "Distributed Optimization and Statistical Learning via the Alternating Direction Method of Multipliers," *North. West. J. Zool.*, vol. 3, no. 1, pp. 1–122, 2011.
[24] Energi Data Serivice, [Online]. Available: https://www.energidataservice.dk/en/dataset/nordpoolmarket